# Strain-Hardening Stages and Structure Evolution in Pure Niobium and Vanadium upon High Pressure Torsion

L. M. Voronova, T. I. Chashchukhina, E. F. Talantsev, M. V. Degtyarev, and T. M. Gapontseva

M. N. Mikheev Institute of Metal Physics, Ural Branch, Russian Academy of Sciences, 18, S. Kovalevskoy St., Ekaterinburg, 620108, Russia

**Abstract.** High pressure torsion (HPT) is one of the ways to form nanostructured materials with high strength properties. However, HPT hardening mechanisms vary from material to material and are poorly understood for some BCC metals, particularly niobium and vanadium. This work aims to identify strain hardening stages for Nb and V metals during HPT. Two approaches have been used to identify the deformation stages during high pressure torsion. The approaches are based on the application of a "piecewise" model, taking into account the different deformation mechanisms that determine the type of the forming structure, and on the analysis of the hardness vs. true strain dependence according to the "$H$–$e^{1/2}$" law. We compared the identified stages with the results of the electron microscopic study of the structure. Both models describe well the structural changes observed microscopically in HPT-deformed niobium. However, we have shown that only the piecewise model gives an adequate description of the stages of structure development in vanadium. We have provided an explanation for the observed difference in the behavior of niobium and vanadium upon HPT.

**Strain-Hardening Stages and Structure Evolution in Pure Niobium and Vanadium upon High Pressure Torsion**

## 1. Introduction

Structural evolution in metals and alloys during plastic deformation can be divided into stages. A deformation stage changes into a new one as a result of a change in the dominant type of dislocation structure, a change in the deformation mechanism (transition to rotational plasticity, activation of twinning) and phase transformation or dynamic recrystallization. Graphically, stages of deformation are represented as sections characterized by different strain hardening laws on the strain curve ($\sigma$–$\varepsilon$). Determining the limit strains at which one structural state changes to another is problematic because the stages on the curve replace each other smoothly (except for the yield point) rather than abruptly at a particular point. Many publications are devoted to the analytical treatment of hardening curves and offer a number of analytical expressions to describe them [1–8]. The hardening curves of BCC metals are usually presented in the form of a parabola [9], but they actually have a more sophisticated shape at higher strains. Some researchers [1, 6–8, 10] identify stages with constant and variable hardening coefficients $\theta_i = \partial\sigma/\partial\varepsilon$, while others [4, 5] identify stages with a constant exponent in $\sigma \sim \varepsilon^n$ (hardening exponent). To identify the different stages of the deformation curve for FCC metals, researchers have used several parameters, including activation enthalpy, dislocation density, and strain rate sensitivity, in addition to the strain hardening coefficient [7, 8]. Up to four stages are typically identified at low strains and up to eight stages are typically identified in the region of severe plastic deformation leading to the formation of a submicrocrystalline structure (SMC) [4, 6–8]. For example, a sequence of eight hardening stages has been identified after deformation of BCC metals: from the first stage, corresponding to a dislocation structure observed at grain boundaries, to a highly misoriented cellular structure, to the stage where new grains form in response to dynamic recrystallization [5]. Identifying the stages on a hardening curve may be necessary for predicting the type of forming structure at a particular strain without the need for comprehensive electron microscopy studies.

To estimate the mechanical properties of nano and submicrocrystalline materials, hardness measurements are often used instead of flow stress measurements [11–13]. In this case, a hardening curve is plotted in hardness-strain coordinates [11,12,14–16]. Various severe plastic deformation (SPD) techniques are employed to form such materials, but the greatest refinement is achieved by high pressure torsion (HPT) deformation [17]. In [18], we identified the stages of structural states in BCC Nb deformed by HPT, assuming that the hardness ($H$) depends on the true strain ($e$), and this dependence is described by a parabolic function. To linearize the parabolic hardness–strain dependence, we used "$H - e^{1/2}$" coordinates [18]. Three linear regions with different slope angles of the $H$–$e^{1/2}$ dependence were observed in the strain range $0.5 < e < 12$. As a result, three stages were identified. According to electron microscopy studies, there is a specific structural state at each stage: dislocation cell structure, SMC structure consisting of highly misoriented microcrystallites, and an intermediate stage of mixed structure. However, firstly, the parabolic law does not work at every stage of deformation [4,5], and secondly, this approach does not take into account the specific deformation mechanisms acting at each stage and affecting the hardening exponent.

A "piecewise model" has been proposed based on analysis of an $H(e)$ dependence [19] to identify the stages of structural states in a material after HPT deformation. A continuous piecewise functional approach underlying the model allows us to describe various complex systems [20]. The model uses true strains (breaking points) as free-fit parameters, which limit the range in which one of the deformation mechanisms operates. The model was applied to the analysis of pure polycrystalline iron, AISI 1020 steel, and AISI 13B20 steel, which allowed us to identify strains, corresponding to the boundaries of structural stages, and to identify the exponents in $H$–$e$ dependence at each stage.

The question of whether this model is universally applicable to other materials remains open. In this work, we apply the "piecewise model" to analyze structural changes in BCC metals (niobium, vanadium) during HPT. The aim of this work is to analyze the hardness vs. true strain dependencies for niobium and vanadium after their room temperature HPT deformation using the above approach (piecewise model), to determine structural stage boundaries and to correlate the results with structural investigations.

**2. Experimental**

Single-crystal niobium (99.95 wt %) and coarse-grained vanadium (99.98 wt %, grain size of several millimeters) were investigated. These metals have the same crystal lattice structure as BCC iron. However, they differ from iron in that they have different stacking fault energy (SFE) values compared to iron. Vanadium has a slightly lower SFE [20], while niobium has a higher SFE [18]. In addition, Nb has a low temperature limit at which the temperature dependence of the flow stress disappears (290 K for niobium vs. 340 K for iron), resulting in abnormally high dislocation mobility in niobium [21, 22]. Several works have shown that the high dislocation mobility associated with the low Peierls stress makes the deformation behavior of niobium and vanadium more similar to that of FCC metals than to that of BCC metals [22–24]. Severe deformation of niobium is reported to result in a steady-state stage, similar to that in nickel, and possible dynamic recrystallization [25, 26] during deformation at room temperature, even though this temperature is only $0.11T_m$. Dynamic recrystallization during HPT deformation develops at temperatures above $0.2T_m$ in pure FCC metals [27]. In niobium, which is characterized by abnormally high dislocation mobility, hardening is delayed, possibly due to dynamic recovery. It should also be noted that there is no twinning or pressure-induced phase transformation in niobium during room temperature deformation. A lower SFE of vanadium increases the probability of twinning [28]. The structural evolution in V prior to the formation of an SMC state is still poorly understood [29–31].

The materials under investigation were deformed using the unconstrained HPT technique [32,33]. Deformation was performed at a pressure of 8 GPa for Nb and 6 GPa for V. The angle of the anvil rotation was varied from 0° (upsetting without shear) to 15x360° (15 revolutions). The samples before deformation were in the form of disks 5 mm in diameter and 0.3 mm thick. As shown in [18], the *H*(*e*) curves for niobium single crystals with {110} and {001} orientations coincided with each other after deformation at 20°C. Therefore, the data for both single crystals are combined in this paper.

The true strain was calculated considering the sample thickness before and after deformation ($h_0$ and $h_{ir}$, respectively), the angle of anvil rotation ($\varphi$), and the distance to the sample center ($r_i$) [16]:

$$e = \ln\left(1 + \left[\frac{\varphi r_i}{h_{ir}}\right]^2\right)^{0.5} + \ln\frac{h_0}{h_{ir}}. \tag{1}$$

The Vickers hardness of the materials was measured by testing along two perpendicular sample diameters with a step of 0.25 mm and a load of 0.5 N. The tests were performed using LOMO PMT-3 (Russia) and Qness GmbH (Austria) hardness testers. Hardness values calculated for different samples of the same material were averaged over true strain intervals of 0.2.

After deformation, the structure of niobium and vanadium was examined using a JEM 200CX transmission electron microscope (Japan, TEM) at an accelerating voltage of 160 kV. TEM was performed at a distance of 1.5 ± 0.2 mm from the center of the sample. Samples for TEM were prepared by mechanical thinning followed by electrolytic polishing. The samples deformed by HPT at an angle of rotation of the anvil of 180° or less were further examined by scanning electron microscopy (SEM) using a Phillips QUANTA200 Pegasus instrument (The Netherlands) equipped with an attachment for analyzing electron backscatter diffraction (EBSD) patterns. EBSD analysis was performed at an accelerating voltage of 20 kV at a scanning step in the range 0.2–2 μm. Points with a confidence index (CI) less than 0.1 and grains containing 4 pixels and smaller were not analyzed.

**3. Piecewise model**

The "piecewise model" has been proposed and described in detail in [19]. The model is based on the assumption of the following postulates:

(1) There are two different deformation mechanisms which are responsible for the type of structure being formed. Each type of structure results in a particular hardness change law for materials undergoing deformation.

For the niobium under study, this means that two deformation mechanisms act during HPT at 20°C, namely, a dislocation mechanism forming a dislocation cellular structure and a rotational mechanism (at higher strains) forming an SMC structure. These two stages have been detected and partially studied in [18]. We assume that the same deformation mechanisms operate in V at room temperature.

(2) There is a true strain region where two mechanisms act simultaneously. The result is a mixed structure of dislocation cells and highly misoriented microcrystallites.

(3) In the true strain range where two types of structures exist simultaneously, the total hardness of a material is calculated using the rule of mixtures. That is, it is the result of a linear summation of the functions identified for each mechanism, taking into account the "weight" of each function at a given point.

4. A simple analytical function describes each of the two deformation mechanisms. In particular, the following functions of hardness vs. true strain are proposed in [19]:

$$\begin{cases} H_1(e) = H_1(0) \times \left(1 + \left(\frac{e}{P}\right)^{\alpha_1}\right) \; stage\ 1 & (2) \\ H_2(e) = H_2(0) \times \left(1 + \left(\frac{e}{PQ}\right)^{\alpha_2}\right) \; stage\ 2 & (3) \end{cases}$$

where $H_1(e)$ and $H_1(e)$ are the hardness functions for structural states 1 and 2, respectively (i.e., in the strain ranges (0, $e1$) and ($e2,\infty$), respectively; $e1$ is the highest strain at which structural state 1 exists and $e2$ is the point at which structural state 2 originates); $P$, $Q$, $\alpha1$, and $\alpha2$ are the free-fitting parameters of the model (Eqs. (2), (3) in [19]). Identical functions were also used in this study.

## 4. Results

### 4.1. Niobium.

Figure 1 shows the results of the "piecewise model" processing of the experimental data obtained for niobium in [18]. Characteristic true strains were found using the model, namely, $e_2 = 2.4 \pm 0.3$, at which microcrystallites begin to form in niobium (rotational deformation mechanism is activated (state 2)), and $e_1 = 5.4 \pm 1.0$, which corresponds to the transition to an SMC structure (translational mechanism stops its dominating and the dislocation cellular structure no longer exists (state 1)).

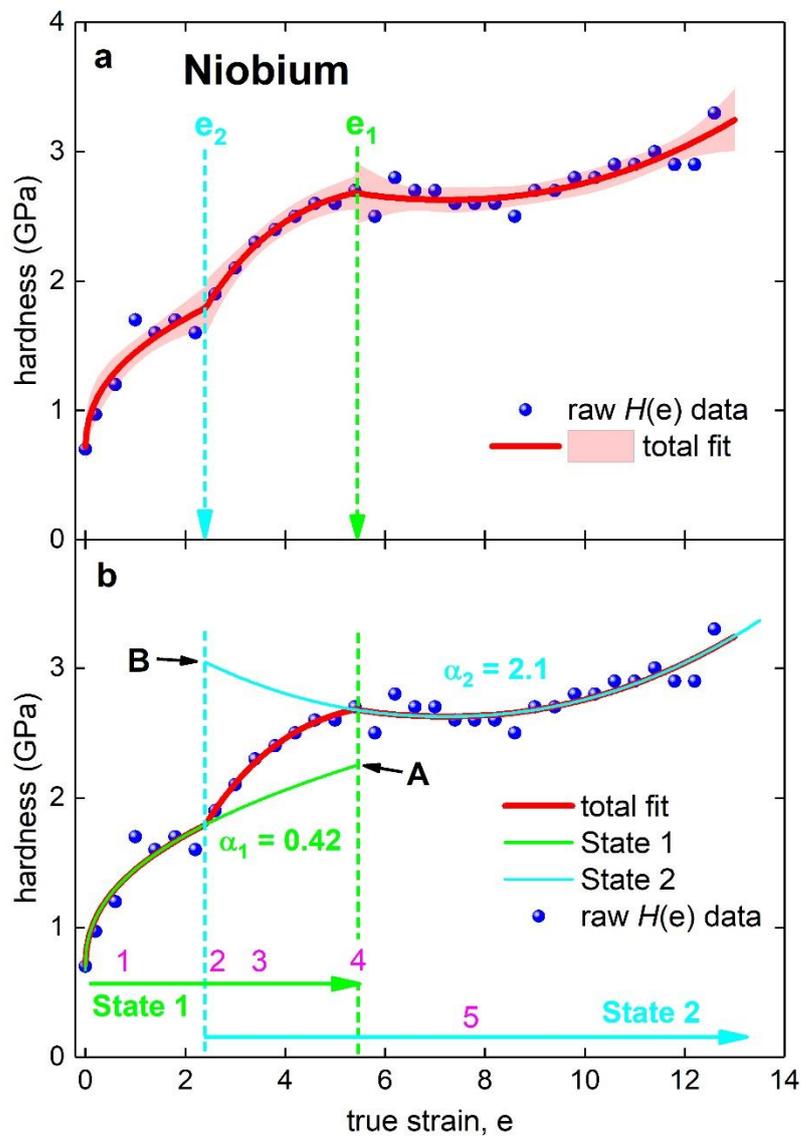

**Fig. 1.** (a) Experimental hardness data ($H(\varepsilon)$) for Nb (points) with a 95% confidence band of hardness, and (b) a "piecewise" model approximation. Contributions of both states are shown in the overlapping strain range; $e1$ and $e2$ are control points, $\alpha1$ and $\alpha2$ are exponents for state 1 and state 2. Numbers *1–5* indicate the true strains at which the niobium microstructure was investigated.

True strain ranges (2.0–2.3) and (3.6–4.0), corresponding to the transition from a cellular structure to a mixed structure and from a mixed structure to an SMC one, respectively, were obtained in the "$H - e^{1/2}$" coordinates in [18]. It can be seen that both the piecewise model and the above dependence give close true strain values at which microcrystallites are expected to form in the structure. However, according to the "piecewise model", an SMC structure forms at a higher strain than that determined by the "$H - e^{1/2}$" dependence. The exponent at the first stage is $α_1 = 0.42$, while at the second stage it is $α_2 = 2.1$.

To verify the model, we analyzed the microstructure of niobium samples with a true strain corresponding to the values indicated by *1–5* in Fig. 1. Figures 2–4 show the evolution of the niobium structure during HPT deformation. We can see a dislocation cellular structure formed after deformation with $e < 2$ (*1* in Fig. 1). The SEM orientation map in the colors of the inverse pole figure (IPF) and the dark field diffraction contrast show a smooth change in the orientation of the initial single crystal (Figs. 2a, 2b). The cell boundaries are dislocation pile-ups (Fig. 2c).

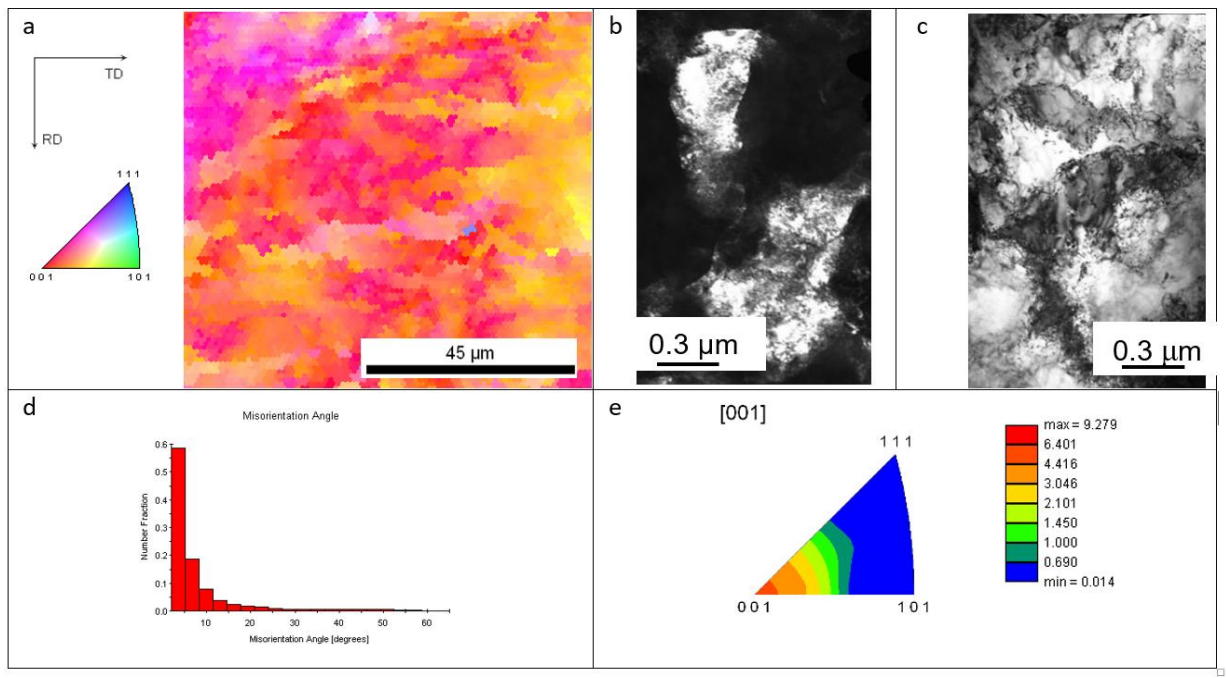

**Fig. 2.** Microstructure of niobium (initial orientation of single crystal {001}) deformed by upsetting with $e$ = 0.6 (*1* in Fig. 1); (a) orientation map in reverse pole figure colors (SEM), (b) dark-field image in the {110} reflection, (c) bright-field image (TEM), (d) distribution of boundaries by angles of misorientation, and (e) IPF.

In the given cellular structure, the average misorientation angle of the cells is 8°. The distribution of angle misorientation is unimodal, with a maximum in the range of small angles up to 5° (Fig. 2d). The IPF shows a blurring of the initial orientation of the single crystal, with a maximum pole density of about 9 (Fig. 2e).

Based on the proposed model, a true strain of 2.3 corresponds to the nucleation of a new structural state (*2* in Fig. 1). EBSD studies (Figs. 3a–3c) indicate that approximately 40% of the high-angle boundaries within the initial single crystal were formed after this deformation. The grain boundary misorientation histogram becomes bimodal: a peak in the region of small angles up to 5°, corresponding to the cellular structure, is retained, and a low, broad peak appears in the region of angles 30°–60° (Fig. 3b).

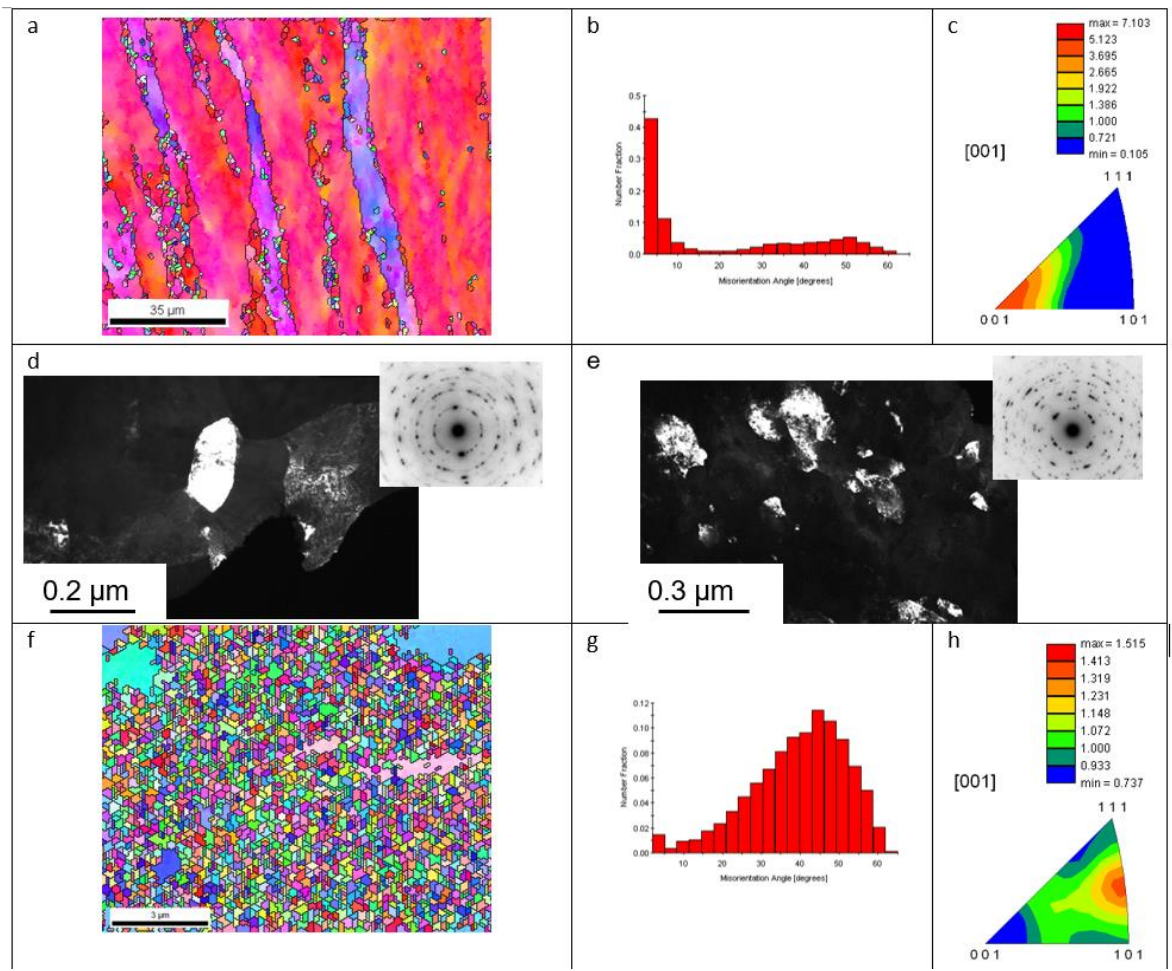

**Fig. 3.** Microstructure of niobium (initial orientation of single crystal {001}) deformed with (a–d) *e* = 2.3 (*2* in Fig. 1) and (e–h) *e* = 3.7 (*3* in Fig. 1); (a), (f) orientation maps in the colors of the IPF with the scheme of high-angle boundaries, (b), (g) distribution of boundaries by angles of misorientation, (c), (h) IPF (SEM), and (d), (e) dark-field images in the {110} reflection (TEM).

The average misorientation angle of the structural elements increases to 20°. Separate microcrystallites (as shown in Fig. 3d) appear as locally misoriented regions on the background of the dislocation cell structure in the dark-field TEM image. The maximum pole density in the IPF decreases, but still corresponds to the orientation of the initial single crystal. The minimum pole density increases by an order of magnitude, indicating an increase in the fraction of other orientations (Fig. 3c). Therefore, a mixed structure with a small fraction of microcrystallites forms at a true strain of $e = 2.3$.

The fraction of microcrystallites in the mixed structure gradually increases with increasing true strain (Figs. 3e–3f). True strain $e = 3.7$ (*3* in Fig. 1) corresponds to a mixed structure with a high fraction of microcrystallites (see Figs. 3e, 3f). The TEM image of the structure shows both microcrystallites and dislocation cells (Fig. 3e). Relatively large areas without high-angle boundaries (HABs) can be seen in the SEM image (Fig. 3f). These are probably regions of cellular structure. The EBSD analysis shows that the fraction of HABs increases up to 96%. The histogram of grain boundary misorientations is similar to that of a random ensemble of grains [34]. A small peak remains in the region of small angles (Fig. 3g). The average misorientation angle of the structural elements reaches 37°. The maximum pole density in the IPF has decreased significantly and does not match the initial single crystal orientation. This indicates a transition from the initial single crystal to a polycrystalline state.

The structure corresponding to $e = 5.6$ (*4* in Fig. 1) is similar in quality to the previous one and differs mainly in the microdiffraction pattern, which shows numerous point reflections producing an almost circular diffraction pattern (Fig. 4a). The structure formed after deformation with $e = 7.9$ (*5* in Fig. 1) is shown in Fig. 4b. It is qualitatively different from the mixed-type structure described above, since it consists entirely of microcrystallites. Locally misoriented, nearly equiaxed microcrystallites are highlighted in the dark-field image. In the electron diffraction pattern, point reflections are uniformly distributed throughout the ring (most clearly seen in the first ring, corresponding to {110} planes). It can be concluded that the SMC structure forms in the deformed niobium after true strain $e = 6.4$ (i.e., the maximum strain at which structural state 1 can exist, according to the piecewise model, $e1 = 5.4 \pm 1.0$).

The electron microscopic study of the microstructure thus confirms that the piecewise model well predicts the deformation ranges in which structural states exist in niobium.

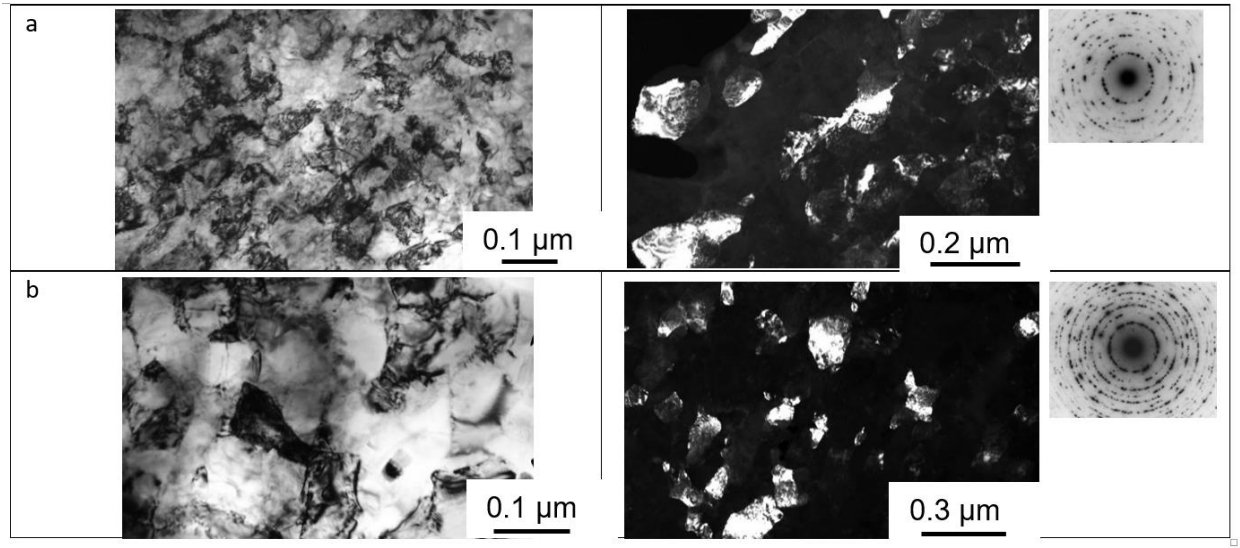

**Fig. 4.** Niobium microstructure (initial single-crystal orientation {001}) after deformation: (a) $e = 5.6$ (*4* in Fig. 1) and (b) $e = 7.9$ (*5* in Fig. 1). Bright-field images are in the left column, and dark-field images taken in a {110} reflection and electron diffraction patterns are in the right column.

Two methods used to identify the stages of the deformation structure ("$H$–$e^{1/2}$", "piecewise" model) gave a good agreement in identifying the strain at which microcrystallites formed in the cellular structure. However, these methods gave a large difference in the strains corresponding to the disappearance of the cellular structure and the transition to the SMC stage. According to the structure studies, the piecewise model gives a more appropriate value of $e1$. This can be attributed to the fact that a small fraction of the cellular structure does not contribute significantly to the hardness, resulting in an apparent mismatch between hardness and structure.

*4.2. Vanadium.*

Figure 5 shows the hardness of vanadium as a function of the true strain. This dependence is shown in the coordinates "$H$–$e$" (Fig. 5a) and "$H$–$e^{1/2}$" (Figs. 5b, 5c). As can be seen, the method previously used to identify the stages of structural states upon deformation of Nb [18] cannot be applied in this case. Figure 5b was constructed assuming that the deformation mechanisms in vanadium are the same as those in niobium and that the same sequence of stages is implemented. However, the lines approximating the second and third stages do not intersect each other in the range of the experimental points. An additional stage between the second and third stages (Fig. 5c) implies the activity of a deformation mechanism other than dislocation and rotation. However, the number of points falling only on this additional approximation line is very small. Assuming that there are four stages of deformation, Fig. 5c gives the following true strain values for the transition from one stage to the next: $e_{1-2} = 1.8$, $e_{2-3}=5.3$, and $e_{3-4}= 9.3$.

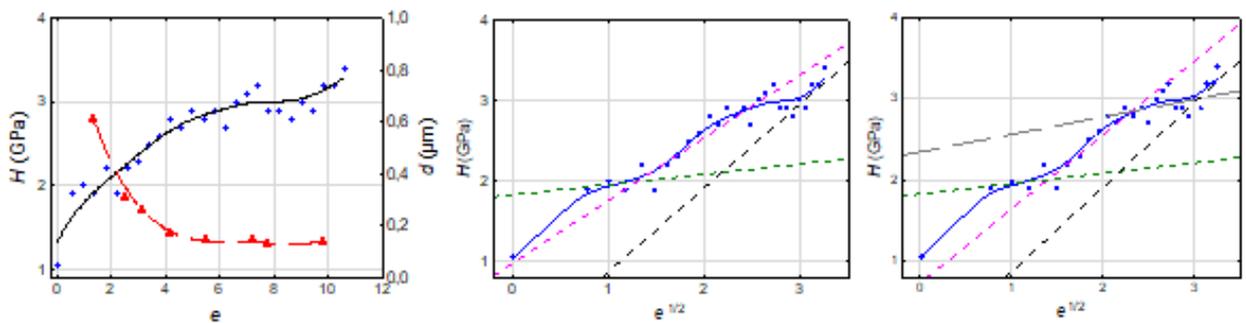

**Fig. 5.** (a) (●) Hardness and (▲) average size of structural elements as functions of the true strain for vanadium; (b,c) identification of stages in the "$H - e^{1/2}$" coordinates.

Figure 6 shows the results of processing the experimental data using the "piecewise model". We can see that there are three stages in vanadium with boundaries at the following true strains: $e2 = 2.8 \pm 0.2$ and $e1 = 7.0 \pm 0.6$. Consequently, the boundaries of the stages of structural states in vanadium, which have been determined by two different methods, are not the same. The type of structure at each stage was identified by electron microscopic study.

Figures 7-9 show the structure evolution in vanadium during HPT deformation. The main distinction between the vanadium structure and the niobium one after upsetting is deformation bands (uncompensated mesobands [35]) formed in the background of the dislocation cellular structure and observed at different scale levels (Fig. 7). The misorientations at the edges of the bands can vary from 15° to 55° (Fig. 7c). The electron diffraction pattern (Fig. 7d) shows that these bands are not twinned with respect to the matrix. As a result, no new possible deformation mechanism, namely twinning, is activated.

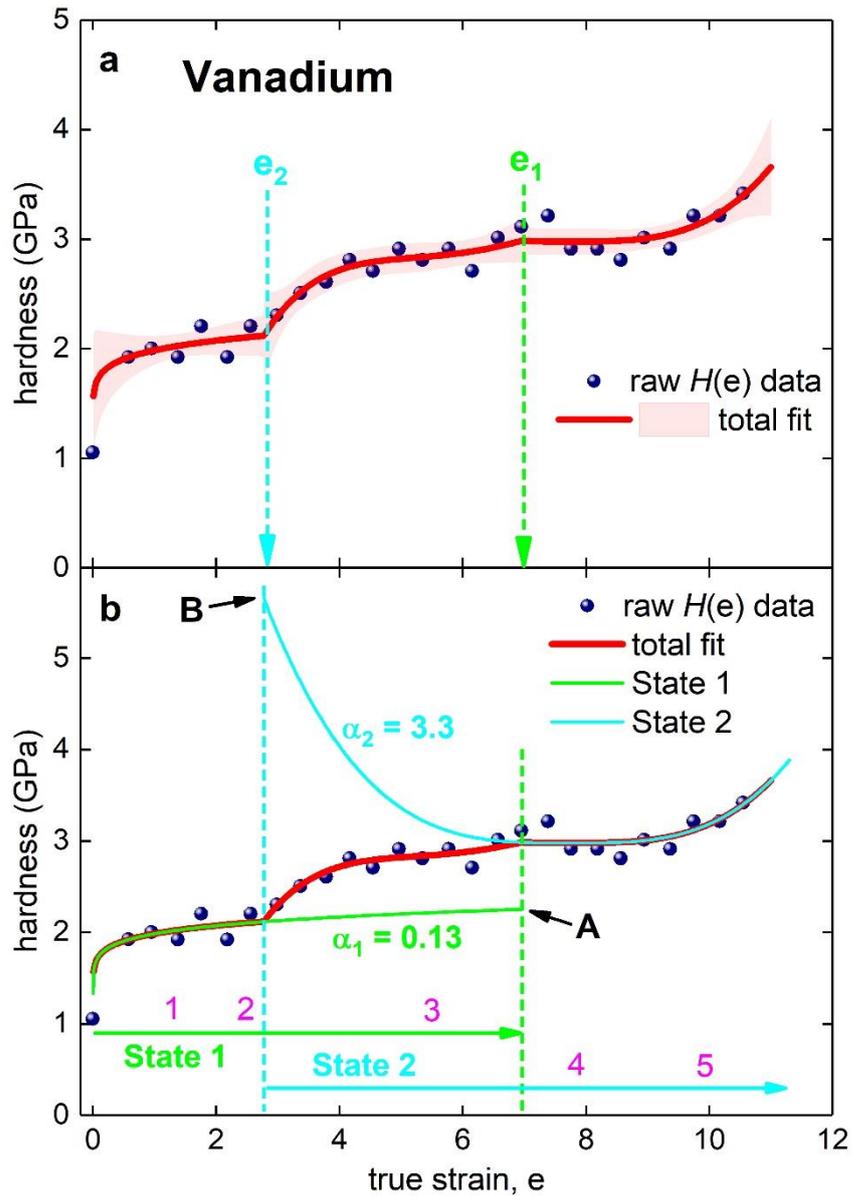

**Рис.6.** (a) Experimental hardness data for V (points) with a 95% confidence band of hardness, and (b) a "piecewise" model approximation. Contributions of both states are shown in the overlapping strain range; $e1$ and $e2$ are control points and exponents, $α1$ and $α2$ are for state 1 and state 2. Numbers *1–5* indicate the true strains at which the vanadium microstructure was investigated.

Deformation bands are also present in the structure after HPT with $e =1.4$ (*1* in Fig. 6). They are fragmented (Figs. 8a, 8b), but no new bands form (Fig. 8c). Microcrystallites were observed after deformation at $e=2.5$ (Fig.9a, corresponding to *2* in Fig.6). The number of microcrystallites in the structure increases and the fraction of cells decreases with increasing true strain (Fig. 9b, corresponding to *3* in Fig. 6). After $e = 7.8$, no dislocation cells are observed in the vanadium structure (Fig. 9c, corresponding to *4* in Fig. 6). Microcrystallites within each deformation band remain closely oriented up to $e=9.8$ (Fig. 9d, corresponding to *5* in Fig. 6). The structural studies show no qualitative changes in the structure at true strains indicated in Fig. 5c ($e_{1-2}$ = 1.8, $e_{2-3}$=5.3, $e_{3-4}$= 9.3). Therefore, the identification of four stages of the structural state is not reasonable.

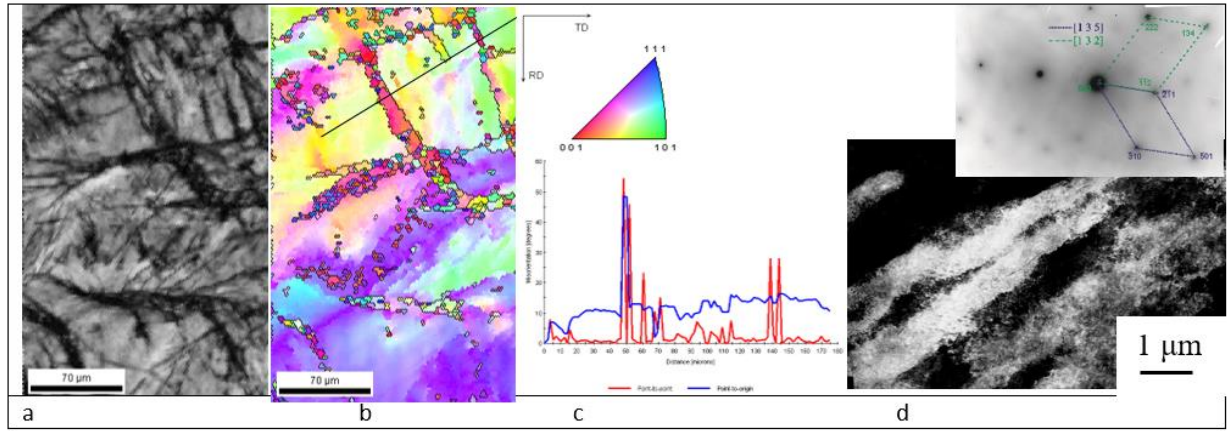

**Fig. 7.** Microstructure of vanadium deformed by upsetting to $e$ = 0.4 in Bridgman anvils: (a) EBSD pattern quality map, (b) EBSD orientation map in the colors of the IPF (with the HAB schematic), (c) the misorientation change along the scanning line shown in Fig. 7b (SEM), and (d) dark-field image in a (2 1 -1) reflection and electron diffraction pattern with [135] and [132] zone axes (TEM).

The fact that the microcrystallites remain similar in orientation up to the maximum true strain achieved indicates that the formation of band structures in the early stages of deformation influences the SMC structure formation during subsequent deformation: turning of adjacent microcrystallites within the initial band is difficult.

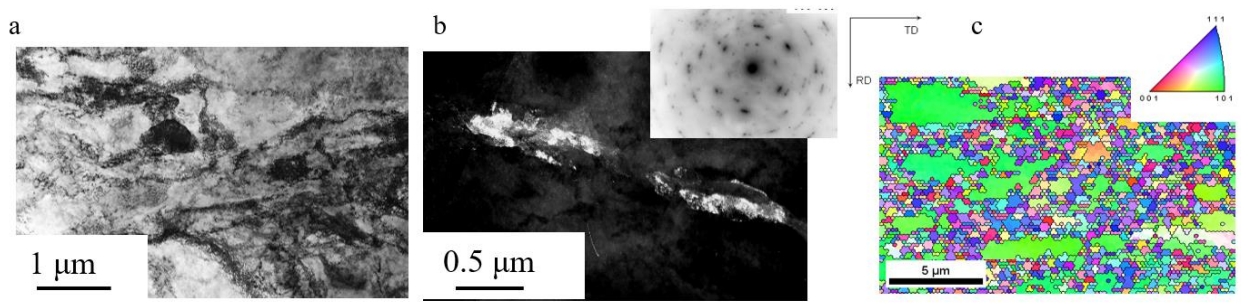

**Fig. 8.** Microstructure of vanadium deformed by HPT to $e$ = 1.4 (*1* in Fig. 6): (a) bright-field and (b) dark-field TEM images and electron diffraction pattern in the insert, and (c) SEM orientation map in the colors of the IPF.

Probably, this structural peculiarity of vanadium did not allow us to identify deformation stages on the "$H–e^{1/2}$" curve. However, the "piecewise" model described adequately the deformation stages in vanadium. The boundaries of the stages of the structural states determined by the model and the structural analysis are practically identical. The first microcrystallites in the structure are detected after $e = 2.5$ (in the model $e2 = 2.8 \pm 0.2$). The cellular structure disappears in the strain range $5.5 < e < 7.8$ (in the model $e1 = 7.0 \pm 0.6$). The average size of the structural elements decreases sharply during deformation up to $e = 4$ and practically does not change after $e > 5$ (Fig. 5a), when a significant number of microcrystallites form in the structure (Fig. 9).

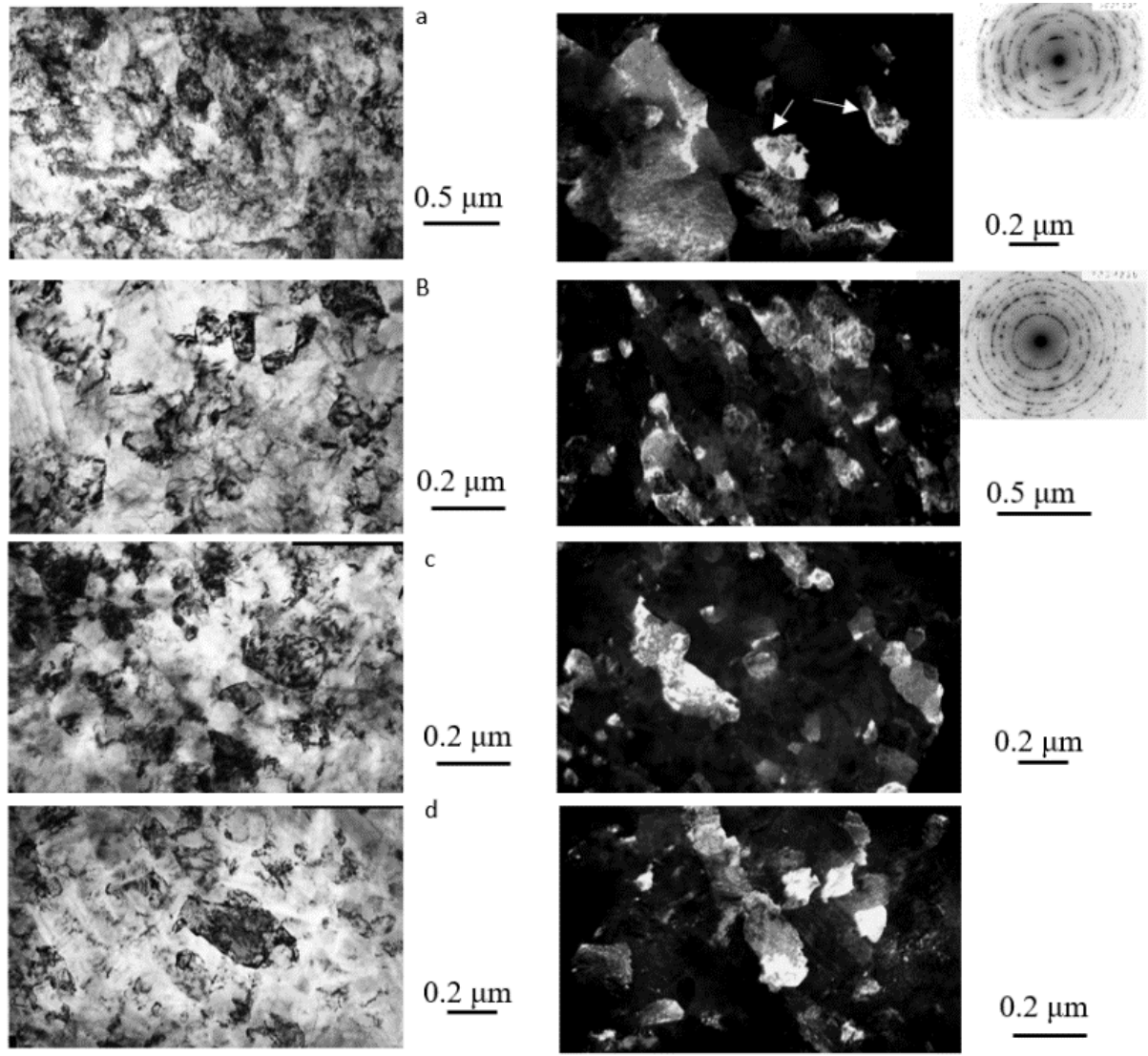

**Fig. 9.** SMC structure in vanadium after HPT: (a) $e = 2.5$ (*2* in Fig. 6), (b) $e = 5.5$ (*3* in Fig. 6), (c) $e = 7.8$ (*4* in Fig. 6), and (d) $e = 9.8$ (*5* in Fig. 6). Bright-field images are in the left column, and dark-field images taken in a {110} reflection and electron diffraction patterns are in the right column. Arrows in Fig. 9a indicate microcrystallites.

## 5. Discussion

The study showed that the piecewise model well describes the structure evolution during HPT deformation of both niobium and vanadium. Moreover, it is the only model that adequately describes the staged evolution of the vanadium structure. Table 1 shows the main parameters of the "piecewise" model for the Nb and V studied in the present work, as well as for pure iron [19]. The strain at which microcrystallites form in the cellular structure ($e2$), i.e., the rotational mechanism of deformation begins to act, is close for all the mentioned materials (see Table 1).

**Table 1.** Parameters of the "piecewise" model characterizing the deformation behavior of pure BCC metals.

| Material / Parameter | Nb | V | Fe |
|---|---|---|---|
| $e1$ | 5.4±1.0 | 7.0±0.6 | 5.4±1.0 |
| $e2$ | 2.4±0.3 | 2.8±0.2 | 3.0±0.1 |
| $α1$ | 0.42±0.11 | 0.13±0.11 | 0.63±0.13 |
| $α2$ | 2.1±1.0 | 3.3±1.3 | 2.1±0.3 |

However, the strain at which the cellular structure is completely replaced by the SMC structure ($e1$), and the rotational mechanism of deformation becomes predominant, is greater in vanadium than in Nb and Fe. The exponent of the $H$–$e$ dependence at the stage of the cellular structure ($α1$) in vanadium is significantly smaller than that in other materials. Consequently, vanadium is characterized by a long stage of simultaneous existence of the cellular and SMC structures and a low hardening intensity in state 1 that corresponds to the cellular structure (Table 1, Fig. 6). This fact is probably related to the formation of band structures at small strains. According to [6], microbanding occurs at the deformation stage characterized by a low hardening coefficient. The formation of a strain band was found in [36] to decrease stresses in a sample, i.e., to decrease strain hardening.

The formation of strain bands in vanadium can be associated with the vacancy–dislocation interaction. A model of the deformation of BCC metals was developed in [37]. It states that there is a deformation temperature $T_0$, below which the interaction of a dislocation with a vacancy is difficult. According to the proposed model $T_0 = 229$–$245$ K for Fe and $T_0 = 396$ K for V. In addition, the calculated energy of vacancy migration in V [37] is much higher than that in Fe (125.4 and 73.3 kJ/mol, respectively). Therefore, dislocation climbing in vanadium during deformation at room temperature (300 K) is difficult because, firstly, the deformation temperature is lower than $T_0$ and, secondly, the vacancy migration energy is high. This makes the formation of a cellular structure difficult at the initial stages of deformation ($e < 1$) and creates conditions for localized deformation, which manifests itself in the formation of band structures. These small strains are almost impossible to achieve during HPT and they correspond to single points on the curve shown in Fig. 6. Therefore, the stage where band structures form due to the action of a separate deformation mechanism cannot be detected in the resulting *H-e* dependence. However, localized deformation affects structure evolution during further deformation, hindering the rotation of structural elements within a band. A true strain required for the transition to the SMC stage was also high for other materials with band structures of various types. For example, this situation was observed for steel with a rack martensite structure [38] and during the cryogenic deformation of iron and nickel, the initial stages of which involved deformation twinning [39, 40].

The study showed that the higher SFE and dislocation mobility in niobium compared to those in iron did not significantly affect the parameters of the "piecewise" model.

The maximum value of the extrapolated hardness curve of vanadium $H(e)$ for state 2, where Eq. (3) is applicable, is $H(e = 2.8) = 5.7$ GPa. This point, according to the piecewise model, corresponds to the deformation at which microcrystallites should appear (this point is marked B in Fig. 6). These high hardness values were not achieved during the experiment. A similar behavior of the extrapolated hardness curves ($H(e)$) for state 2 was also observed for niobium (Fig. 1), pure iron (Fig. 3(b) in [19]), and steels (Figs. 6, 7 in [19]). It's noteworthy that the value at point B in Fig. 6 is similar to the theoretical shear strength of vanadium (5 GPa) that was estimated using pseudopotential electronic structure methods [41]. The highest hardness of vanadium, which can be achieved due to the maximum refinement of the cellular structure, is at point A (Fig. 6). This hardness is much lower than the hardness at point B. This relationship is also true for niobium (Fig. 1).

## 6. Conclusions

This work showed that the piecewise model can be used to analyze the strain-hardening stages and structural evolution of BCC niobium and vanadium metals during HPT.

The piecewise model revealed typical true strain $e_2$, at which the submicrocrystalline structure begins to form (the rotational mechanism of deformation is activated), and $e_1$, which corresponds to the transition to the SMC structure (the rotational mechanism becomes dominant). For niobium $e_2 = 2.4 \pm 0.3$ and $e_1 = 5.4 \pm 1.0$, and for vanadium $e_2 = 2.8 \pm 0.2$ and $e_1 = 7.0 \pm 0.6$. The values found are in good agreement with the results of the structural study.

In niobium with high SFE and dislocation mobility, the piecewise model yielded the following exponents that characterize the level of strain hardening: $0.42 \pm 0.11$ in state 1, where the dislocation deformation mechanism is at work, and $2.1 \pm 1.0$ in state 2, where the rotational deformation mechanism is at work.

In vanadium, deformation is localized in the early stages of HPT ($e < 1$). As a result, band structures are formed. They do not affect the formation of the first microcrystallites, but contribute to the transition to the SMC structure at higher true strain. Consequently, vanadium is characterized by a prolonged stage of coexisting cellular and SMC structures. The band structures provide a low intensity of strain hardening in state 1, the state corresponding to the cellular structure (according to the "piecewise" model, exponent is $0.13 \pm 0.11$).

**Acknowledgments**

The authors thank the financial support provided by the Ministry of Science and Higher Education of Russia (theme "Pressure" No. 122021000032-5).